\documentclass[aps]{revtex4}

\usepackage{amsmath,amssymb}

\usepackage{epsfig}
\usepackage{graphicx}

\newcommand{\nn}{\nonumber}

\newcommand{\beq}{\begin{equation}}
\newcommand{\eeq}{\end{equation}}
\newcommand{\ba}{\begin{eqnarray}}
\newcommand{\ea}{\end{eqnarray}}
\newcommand{\ov } {\over }
\newcommand{\p }{\partial }

\def\lb{\lambda }
\def\a{\alpha }
\def\k{\kappa }

\begin{document}

\title{Exact solution of scalar field cosmology with  exponential
  potentials \\
and transient acceleration}

\author{Jorge G. Russo }
%\footnote{jrusso@ecm.ub.es} }

\date{\today}

%\address{~}

\affiliation{Instituci\' o Catalana de Recerca i Estudis
  Avan\c{c}ats (ICREA), \\
  Departament ECM,
Facultat de F\'\i sica, Universitat de Barcelona, Spain}

\begin{abstract}

We show that the general solution of 
scalar field cosmology in $d$ dimensions with  exponential potentials
for  flat Robertson-Walker metric
can be found in a straightforward way by introducing new variables
which completely decouple the system.
The explicit solution shows the region of parameters where the
expansion has
eternal acceleration, transient periods of acceleration, or no period
of acceleration at all. 
In the cases of transient acceleration,
the energy density exhibits a plateau during the accelerated
expansion, where $p\cong -\rho$, due to dominance of potential energy.
We determine the interval of accelerated expansion in terms of a
simple formula. In particular, it shows that
the period of accelerated expansion  decreases in higher dimensions.

\end{abstract}


%; \, \hskip 7 cm hep-th/0403010}

\maketitle

Understanding the origin of dark
energy in the context of fundamental theories such as M/string theory
became an important problem, after the recent WMAP data
\cite{wmap} confirming
that  nearly $70\% $ of the energy density of the universe is in the form
of dark matter.
The dimensional reduction of these fundamental theories to four dimensions
typically gives rise to scalar
fields with exponential potentials coupled to four-dimensional gravity.
Given that a substantial fraction of the energy density of the universe might
indeed consist of quintessence \cite{perl} 
in the form of a slowly-rolling scalar field,
it is of interest to understand whether exponential potentials
could describe observational data for the late-time cosmic acceleration.

Exponential potentials in four dimensions were much investigated in the past
(see e.g. \cite{exponencial}), and several exact solutions
have appeared
(for a recent discussion and further references, see \cite{kehagias}).
In particular, a general solution  in four dimensions was 
obtained in \cite{chimento}
and more recently in \cite{townsend}, and previous discussions of
solutions in $d$ dimensions can be found in \cite{chim2}.
Here we will show that the simplest case of a homogeneous scalar field
coupled to an exponential potential can be actually solved in a direct
and straightforward way in $d$ dimensions by the introduction of 
new variables which
 decouple the system. 
The resulting general solution expressed
in a suitable time frame is remarkably simple. 
The presence of exponential potentials  in 
higher dimensional theories obtained from fundamental theories
is quite generic, so
the $d$ dimensional case is of particular interest 
to compare cosmology
in different dimensions, 
and to see to what extent some physical properties are generic or 
peculiar to four dimensions.

% Having
% scalar field cosmologies that can be solved exactly in terms
% of closed analytic expressions can provide
% a deeper insight in the physics of the model and the role of its parameters.

For four-dimensional cosmologies with  exponential potentials,
some new interesting aspects were recently  pointed out in \cite{town}.
There it was shown
an example of cosmology 
which starts with a decelerating expansion, at some point
it experiences a transitory  period of
acceleration, and it ends with decelerating expansion again.
The origin of the acceleration was further clarified  in \cite{empa}
(related discussions and generalizations can be found in \cite{recent}).
The explicit general solution presented here gives a clear view of the 
window of
parametric space where there is such period of  accelerated expansion,
and the dependence of this period on the dimensionality of the space.

%Finally, we  present the generalization to $d$ dimensions by applying
%the same method to decouple the equations, and  
%describe some features of the solutions.

\bigskip

We start with the action for 4d Einstein gravity coupled to a scalar
field with an exponential  potential $V$:
\beq
S=\int d^4x\sqrt{-g}\bigg({R\over 2\kappa^2} -
{1\ov 2} g^{\mu\nu} \p_\mu \phi \p_\nu\phi
-V_0 \exp(-\lambda \phi )\bigg)
\eeq
We will work in units where $\k^2=8\pi G_N=1/2$. Here $V_0>0 $, 
$\lb >0$  (the case $\lb <0 $ is connected to the case $\lb >0$ by 
the change $\phi\to -\phi $).

Consider homogeneous and isotropic cosmologies described by 
the Robertson-Walker metric for a flat universe ($k=0$)
$ds^2=-dt^2+a^2(t ) dx^i dx^i  ,\ \  i=1,2,3\ .
$
The action takes the form
\beq
S=\int d^3x dt \bigg( - 6 a \dot a^2+ 
a^3\big( {1\ov 2}\dot \phi ^2 -  V_0 \exp(-\lambda \phi ) \big)\bigg)\ .
\eeq
The equations of motion are
\ba
&&\ddot \phi + 3H\dot \phi  - \lambda V_0 e^{-\lambda \phi }=0\ ,
\\
&&H^2={1\over 6}\big({1\ov 2}\dot \phi ^2 +V_0 \exp(-\lambda \phi ) \big)\ ,\ \ \ \ 
 H={\dot a\ov a}\ .
\ea
{}For $V_0>0$, they have the late-time attractor solution:
\ba
&&\phi={2\ov \lb } \log ({t\ov t_0})\ ,\ \ \ a=a_0\big( {t\over
 t_0}\big)^p,\ \ \ p={1\over \lb^2}\ ,
\nn\\
&& \lb^2 V_0 t_0^2=2(3p-1)\ .
\label{attra}
\ea
The solution exists provided $3p>1$, i.e. $\lb< \sqrt{3} $.
The solution (\ref{attra}) shows that the expansion exhibits eternal
acceleration when $\lb < 1$. 
% It was obserbed in [town] that for exponential
% potentials the expansion may also exhibit transient periods of acceleration.
% {}From the explicit expression of the 
% general solution presented here, it  will be clear
% which are the solutions that have this property.

Now we perform a change of variables $\{ a(t),\phi (t) \}\to \{ u(t), v(t) \} $
defined as follows:
\beq
\phi={1\over\kappa }\sqrt{ 2\over 3 }(v-u)\ ,\ \ \ \ 
a^3=e^{v+u}\ .
\eeq
The action takes the form
\ba
&&S=-\int d^3x dt\  e^{u+v}
\bigg(  {8\over  3}\dot u \dot v +V_0e^{-2\a (v-u)}\bigg)
 \ ,
\label{uno}
\\
&&\a \equiv  {\lambda\over  \sqrt{3}}\ .\nn 
\ea
Next, we introduce a new time coordinate $\tau $:
\beq
{d\tau \over dt}= \sqrt{3V_0\over 8} \ e^{\a (u-v)}\ .
\eeq
The action becomes
\beq
S=-  \sqrt{8V_0\ov 3} \int d^3x d\tau  \ e^{u+v} e^{\a (u-v) } \big( u'  v' +1 \big)
\ ,
\eeq
where prime denotes derivative with respect to $\tau $.
The equations of motion become decoupled,
\ba
v''+(1-\a ){v'}^2-1-\a &=& 0\ , \nn\\   
u''+(1+\a ){u'}^2-1+\a &=& 0\ , 
\label{dos}
\ea
and  the Hamiltonian constraint reduces to
\beq
v'u'=1\ .
\eeq
Let us first consider the special case $\a=1$. 
This corresponds to the limiting value for
the attractor solution $\lb = \sqrt{3}$. 
 The direct integration of (\ref{dos}) gives
\beq
u={1\ov 2} \log (2\tau)\ ,\ \ \ \ v=\tau ^2\ ,
\eeq
where the integration constants
have been removed by rescaling the coordinates 
$x_i$, by a shift in time and by a shift in the scalar field.
Thus
\ba
ds^2 &=& {8\ov 3V_0} {e^{2\tau^2}\ov 2\tau }d\tau^2+
e^{2\tau^2\ov 3} (2\tau)^{1\ov 3}dx^i dx^i
\ ,\ \nn \\
\phi &=&
{1\over \sqrt {3}} \big( 2\tau^2 -\log (2\tau) \big)\ .
\ea
At late times, this has the behavior $a\sim t^{1/3}$ and $\phi\sim
{2\ov \sqrt{3}}\log t$, which coincides with the 
attractor solution in the  limit $\lambda= \sqrt{3}$.

The case $\a=1/2$
can be studied directly in terms of 
the original time coordinate $t$, by the action (\ref{uno}),
 introducing variables $X=e^u$, $Y=e^v$. Then we have the equations 
$\ddot X=0$, $\ddot Y={3\over 4}V_0 X$,
 which are readily solved. 
%The solution is
% \beq
% X= a_1 t\ ,\ \ \ \ Y={1\over 8}V_0 a_1 t^3 + c_1 \ .
% \eeq
 This leads to
\beq
a^3 = {1\over 8}V_0  t^4 + c_1 t\ ,\ \ \ 
\phi ={2\over \sqrt {3}} \log \big({1\over 8}V_0  t^2 + {c_1\over  t}
\big)\ .
\eeq
%The same solution follows of course from the system (\ref{dos}).

Let us now consider the general solution to the system (\ref{dos}).
It is convenient to discuss separately the cases $\a <1$ and $\a >1 $.
\medskip

\noindent a) Case $\a<1$ ($\lb<\sqrt{3}$):
\beq
u=\sqrt{1-\a \over 1+\a }\ \tau +
{1\over 1+\a }\log \big( 1- me^{-2w\tau } \big)
\label{xxu}
\eeq
\beq
v=\sqrt{1+\a \over 1-\a }\ \tau +
{1\over 1-\a }\log \big( 1+ me^{-2w\tau } \big)
\label{xxv}
\eeq
with 
$w\equiv \sqrt{1-\a^2}$.
Thus the metric and scalar field are
\ba
ds^2 &=& -{8\over 3V_0} e^{4\a ^2 \tau\ov w} 
{(1+m e^{-2w\tau } )^{2\a\ov(1-\a)}
\over (1-m e^{-2w \tau} )^{2\a\ov(1+\a)}}  d\tau ^2 +
 e^{4 \tau\ov 3w} (1+m e^{-2w\tau } )^{2\ov 3(1-\a)}
 (1-m e^{-2w \tau} )^{2\ov 3(1+\a)}dx^i dx^i \nn
\\
\phi &=&{2\over \sqrt{3}} \bigg( {2\a \tau \ov w}-
{1\over 1+\a}
 \log(1-m e^{-2w\tau } ) 
+ {1\over 1-\a }\log(1+m e^{-2w \tau })\bigg)\ .
\label{meti}
\ea
{}For $m\neq 0$ 
the  absolute value of $m$ can be absorbed
into a shift of $\tau$.
The attractor solution is the particular case $m=0$, and it is the asymptotic
limit of the whole family of solutions.
{} Indeed, for large $\tau $, we have $t=\sqrt{8\ov 3V_0}{w\ov 2\a^2}
\exp (2\a^2\tau/ w)$. Substituting
into the scalar field $\phi $ and into the scale factor $a^2$, we reproduce
the 
attractor solution (\ref{attra}). 
%In general, the change of time coordinate from $\tau $ to proper time $t$ 
%involves hypergeometric functions.

Note that the solution (\ref{meti})
also includes the case when $\lb =0$, corresponding to a cosmological constant.
In this case $\tau =\sqrt{3V_0\ov 8}\ t$, 
and the general solution is:
$$
a^3=a_0^3e^{2 \tau  }(1 -m^2 e^{-4 \tau })\ ,\ \ \ \ 
\phi={2\over \sqrt{3}}\log 
( {1+me^{-2 \tau }\over 1-m  e^{-2 \tau } })
$$
When $m=0$, we recover  the standard de Sitter solution with 
$\phi=$constant and $a=a_0e^{2\tau/3}=a_0e^{  \sqrt{V_0\ov 6} t    } $ .

%The solutions discussed in these papers can 
%be related to the $d=4$, $\a <1$ case discussed
%here by analytic continuation of $n$ 
%to continuous (negative or positive) values. 

\medskip

\noindent b) Case $\a>1$ ($\lb>\sqrt{3}$): 

The general solution is given by
\beq
u={1\ov \a +1 } \log \big[\sin \beta\big]
\ ,\ \ \ \ \ 
v=-{1\ov \a -1} \log \big[\cos \beta\big]
\label{betxx}
\eeq
where $\beta\equiv\sqrt{\a^2-1}\ \tau +\beta_0$. Thus 
 \ba
ds^2 &=& -{8\over 3V_0} (\cos \beta )^{-{2\a\ov \a-1}} 
(\sin\beta )^{-{2\a\over \a+1}}  d\tau ^2 + {(\sin \beta )^{{2\over 3(\a+1)}} \over
(\cos \beta )^{{2\over 3(\a-1)}} }dx^i dx^i\ ,
\label{betyy}
\\
\phi&=&-{2\over \sqrt{3}} \bigg({1\over \a +1 }
 \log(\sin  \beta )+ {1\over \a -1 }\log(\cos \beta )\bigg).\nn
\ea

\medskip

Let us now consider the equation of state at early and late times.
We first consider the case $m>0$ and $\a<1 $.
Setting $m=1$, 
time $t\cong 0$ corresponds to $\tau $ with $1\cong e^{-2 w\tau}$.
In the vicinity of this point, eqs. (\ref{meti}) imply 
\beq
t\sim \tau^{1\ov \a +1}\ ,\ \ \  a^3\sim \tau^{1\ov \a +1}\ ,\ \ \ 
\phi=-{2\ov \sqrt{3} (1+\a)}\log 2w\tau\ ,
\label{qqw}
\eeq
i.e. $a\sim t^{1/3} \ .$
This corresponds to the kinetic attractor, with the equation of state
$p=\rho $ corresponding to matter with dominance of kinetic energy. 
The same result $a\sim t^{1/3}$ 
holds for $m<0$. This is
clear from eqs. (\ref{xxu}), (\ref{xxv}) since the change 
$m\to -m$ is equivalent to
changing $\a\to -\a$ and $u\to v$, which leaves  the metric invariant.
For large $t$, one has the standard
result $a\sim t^p$ that follows from the late-time attractor
solution,
corresponding to $p=\omega \rho $, $\omega= 2\a -1={2\ov 3}\lb^2-1$.

The cases that arise from dimensional reduction to four dimensions
on  hyperbolic space ${\cal H}_n$ (discussed in
\cite{town,empa,recent}) correspond to a discrete set of $\lambda $
of the form $\lambda=\sqrt{(n+2)/n},$ with $n= 2,3...$~. 
In these
cases, the asymptotic behavior $a\sim t^p$ has a simple interpretation 
as arising from higher dimensional Milne space \cite{russo}.

Fig. 1 shows a plot of the energy density as a function of $\tau $
for $\a =2/3$ and $m>0$, which exhibits the generic behavior 
for any  $\a $ in the interval $(1/\sqrt{3},1)$.
 The kinetic energy vanishes at a certain time; this is the
instant where $\phi $ is reflected off the potential barrier, as
observed in \cite{empa}.
We see that the energy density 
exhibits a plateau during the period of acceleration,
in agreement with the fact that near $\dot \phi=0$ the equation of
state is $p\cong -\rho $ so that $\dot \rho=-3H(\rho+p)\cong 0$.
%The period of acceleration can be appreciated in fig.~2. 
The potential as a function of time intersects the curve
representing
twice the kinetic energy at two points.
Since one of the Einstein equations imply that 
$R_{00}$ is proportional to $2T-V$, the period between these
two points is precisely the period 
of accelerated expansion.

\begin{figure}[h!]
\centering
\includegraphics*[width=169pt, height=130pt]{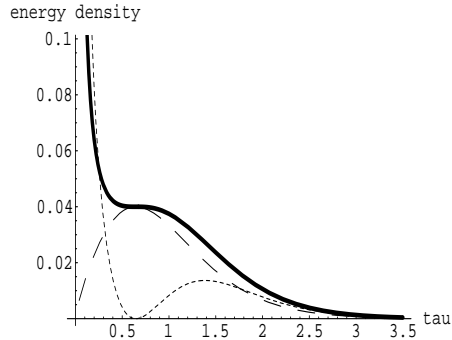}%
\caption{Energy density as a function of time $\tau $ for $\a=2/3$ 
(solid line). The period of accelerated expansion can be visualized as
the interval between the intersection of twice the kinetic 
energy (short-dashed line) and  the potential energy (long-dashed
line). Here  $\a=2/3$ and the same qualitative behavior
appears for all $\a \in (1/\sqrt{3},1)$, $m>0$.}
\end{figure}
%

%
%\begin{figure}[h!]
%\centering
%\includegraphics*[width=169pt, height=130pt]{ttvv.eps}%
%\caption{The period of accelerated expansion can be visualized as
%the interval between the intersection of twice the kinetic 
% energy (short-dashed line) and  the potential energy (long-dashed
%line). Here  $\a=2/3$ and the same qualitative behavior
%appears for all $\a \in (1/\sqrt{3},1)$, $m>0$.
%}
%\end{figure}
%

Now consider the case $\a > 1$, where the solutions are given by 
eq.~(\ref{betyy}).
The geometry has a singularity at $\beta=0$ and $\beta={\pi\ov
  2}$. Setting $\beta_0=0$ by a shift in $\tau $, the time between
the two singularities corresponds to
the interval $\tau\in \big( 0,\pi/(2\sqrt{\a^2-1})\big)$. By integrating
${dt\ov d\tau}$, we see that this in turn corresponds to a {\it infinite}
proper time interval $t\in (0,\infty)$.
In the vicinity of  the two singularities, the geometry has the same behavior
(\ref{qqw}).
Computing $\dot a$ one finds that it is positive definite, so the
geometry describes an expanding universe, which starts
and ends with a decelerating expansion.

{}For $\ddot a$ we
find an equation similar to (\ref{asdx}), 
with $Z\equiv \cos(2 \sqrt{\a^2-1})\tau )$. The two roots are
given by eq.~(\ref{aqua}).
%\beqZ_{1,2}={2\a \pm \sqrt{3}(\a^2-1) \over 3\a^2-1}\ .
%\label{aqu}\eeq
{}For $\a>1$ one has $|Z_{1,2}|<1$, so that the equation 
$Z_{1,2}\equiv \cos(2 \sqrt{\a^2-1})\tau_{1,2} )$ always has  real
solutions $\tau_{1,2}$ in the interval 
$0<\sqrt{\a^2-1}\tau_{1,2}< {\pi\ov 2}$, with $\tau_1(\a)<\tau_2(\a
)$. 
As a result, $\ddot a$ is positive
in the interval between $\tau_1$ and $\tau_2$. Thus solutions with
$\a>1 $ also exhibit a transitory period of acceleration.

We now consider the generalization to $d$ dimensions:
\beq
S=\int d^dx\sqrt{-g}\bigg({R\over 2\kappa^2} -{1\ov 2} g^{\mu\nu} \p_\mu \phi \p_\nu\phi
-V(\phi )\bigg)
\eeq
Setting $\k^2=8\pi G_N=1/2$, and considering again
 the Robertson-Walker metric for a flat universe ($k=0$)
\beq
ds^2=-dt^2+a^2(t ) dx^i dx^i \ ,\ \ \ \ i=1,...,d-1
\eeq
the action  takes the form
\beq
S=\int d^{d}x  \bigg(-(d-1)(d-2) a^{d-3} \dot a^2+ a^{d-1}
\big( {1\ov 2}\dot \phi ^2 -V(\phi ) \big)\bigg)\nn
\eeq
Now we need to introduce the analogous $u,v$ variables. 
They are defined as follows:
\beq
\phi={1\over\kappa }\sqrt{ d-2\over d-1 }(v-u)\ ,\ \ \ \ 
a^{d-1}=e^{v+u}
\label{qqa}
\eeq
The action becomes
\ba
S &=& -\int d^{d-1}x dt \ e^{u+v}\bigg(  {2 (d-2)\over \kappa^2 (d-1)}\dot
u \dot v +V_0e^{-2\a (v-u ) }\bigg) ,\nn
\\
\a &\equiv &
 {1\over 2\kappa }\sqrt{ d-2\over d-1 }\ \lambda \ .
\ea
Introduce a new time variable $\tau $
\beq
{d\tau \over dt}=\kappa  \sqrt{(d-1)V_0\over 2(d-2)} \ e^{\a (u-v) }
\label{xxx}
\eeq
The action becomes
\beq
S=- {1\ov \kappa}\sqrt{2(d-2)V_0\over d-1} 
\int d^{d-1}x d\tau \ e^{u+v} e^{\a (u-v) } \big( u'  v' +1 \big)
\ .\ \ \ 
\eeq
Thus we get the same system for all dimensions. The solutions for 
$u$ and $v$ are
the same as before. In particular, the general solution
is given by eqs. (\ref{xxu}), (\ref{xxv}) for $\a<1$, and
by eqs. (\ref{betxx})  for $\a>1$.
The metric and the scalar field are then read from 
(\ref{qqa}) and (\ref{xxx}). For $\a<1$:
\ba
ds^2 &=& -{2(d-2)\ov \kappa^2(d-1)V_0}\ e^{4\a ^2 \tau\ov w} 
{(1+m e^{-2w\tau } )^{2\a\ov(1-\a)}
\over (1-m e^{-2w \tau} )^{2\a\ov(1+\a)}}  d\tau ^2 
+ e^{4 \tau\ov sw} (1+m e^{-2w\tau } )^{2\ov s(1-\a)}
 (1-m e^{-2w \tau} )^{2\ov s(1+\a)}dx^i dx^i \nn\\
\phi &=&{1\over \kappa} \sqrt{d-2\ov d-1}
\bigg( {2\a \tau \ov w}-
{1\over 1+\a}
 \log(1-m e^{-2w\tau } ) + {1\over 1-\a }\log(1+m e^{-2w \tau })\bigg)\ ,
\label{metid}
\ea
with $s=d-1$.
The solution has the following behavior:
\ba
&&a \sim t^{1\ov d-1}\ ,\ \ \ \ \phi=-{1\ov \kappa}\sqrt{d-2\ov
  d-1}\log t\ ,\ \ {\rm for}\ \ t\cong 0\ ,\nn
\\
&&a \sim t^{4\kappa^2\ov (d-2)\lb^2}\ ,\ 
\  \ \ \phi={2\ov \lb}\log t\ ,\ \ \ \ 
\ {\rm for}\ \ t\gg 1\ .
\label{dfg}
\ea
The equation of state at initial times is $p=\rho$, while at late
times 
\ba
p=\omega \rho , \ \ \ \  \omega={d-2\ov 2\k^2(d-1)}\lb^2-1\ .
\ea
The geometry describes an expanding cosmology. According to
eq. (\ref{dfg}),
the expansion has eternal acceleration if $\lb < 
{2\kappa \ov  \sqrt{d-2} }$.
We now show that when ${2\kappa\ov  \sqrt{d-2}} < 
\lb < 2\kappa \sqrt{d-1\ov d-2}$, 
the solution (\ref{metid}) with $m>0$ has a transient period
of acceleration, just as the $d=4$ solution.
Computing ${da\over dt}={da\ov d\tau } {d\tau\ov dt}$ from
(\ref{meti}), one sees that $\dot a$ is proportional to a quantity
which is positive definite for all $m$ and $\tau $. 
Setting $|m|=1$ by a shift of $\tau $ and computing $\ddot a$, we obtain:
\ba
\ddot a =-({\rm positive}) \bigg( \big( (d-1)\a^2-1\big) Z^2
-2(d-2)\, {\rm sign}(m) \a Z+ d-1-\a^2\bigg) 
\label{asdx}
\ea
with $Z\equiv \cosh(2w\tau )$.
If $(d-1)\a^2<1$ (corresponding to $\lb< {2\kappa \ov  \sqrt{d-2} }  $), 
we get eternal acceleration for all $m$, since at late times
the first term dominates.
This is expected, in
consistency with the attractor solution. 
If $(d-1)\a^2>1$, at late times
the solution always describes a decelerating expansion. 
If  $(d-1)\a^2>1$ and
$m<0$,
then the right hand side of (\ref{asdx}) is negative definite, implying
deceleration at all times. Finally, in the case $(d-1)\a^2>1$ and
$m>0$, the solution always exhibits the  transient periods of
acceleration
noticed in \cite{town}. 
Indeed, eq.~(\ref{asdx}) has two roots
\beq
Z_\pm={(d-2)\a \pm \sqrt{d-1}(1-\a^2) \over (d-1)\a^2-1}\ .
\label{aqua}
\eeq
This determines the interval   $\tau_-(\a)<\tau <\tau_+(\a )$  
of accelerated expansion.
The roots $\tau_{\pm }$ are real (and positive), since in 
the relevant interval 
$\a \in (1/\sqrt{d-1},1)$ we have $Z_{\pm }>1$.
In the limit $\a\to 1$, the two roots coincide, 
and the period of accelerated expansion goes to zero.
In the opposite limit, $\a\to 1/\sqrt{d-1}$, one gets
$Z_-=d/(2\sqrt{d-1})$ and $Z_+\to\infty $, so there is an infinite
period
of acceleration. 
{}For large dimensions, one gets $Z_\pm =1/\a \pm
(1-\a^2)/(\a^2\sqrt{d})+O(1/d)$,
which shows that the transient period of acceleration is shorter in
higher dimensions.

% The  existence of a period of acceleration
% also agrees with the discussion of \cite{empa}.

In conclusion, 
we have shown that scalar field cosmology with
an  exponential
potential can be solved in a simple way in any spacetime
dimension by the introduction of
 suitable variables which completely decouple the system.
We have given the explicit solution in $d$ dimensions, which  exhibits
 initial and late time attractors, and determine the corresponding
equations of state.
The mechanism of transient acceleration observed in \cite{town}\ 
subsists in higher dimensions. This could be of relevance for
string/M-theory cosmological scenarios for early universe where $d$ 
dimensions are comparable and much larger than the Planck length, so that the
 effective field theory description applies.
{}In four dimensions, as discussed in \cite{townsend}, 
the effect of transient acceleration could
explain the observed accelerated expansion 
of the universe, but it seems insufficient to
model early universe inflation. However,
as pointed out above, the period of accelerated expansion can be very
 large for exponential potentials with $\alpha\sim 1/\sqrt{3}$. 
This  effect could also apply
in cosmologies with more general potentials exhibiting periods of
 transient
acceleration, i.e. there could be critical values of the parameters where
the acceleration period tends to infinity. In order to describe
inflation within an effective field theory approach,
it seems worthwhile to search for string theory compactifications 
that could give rise to such potentials which are close to these
critical potentials.

\smallskip

{\it Acknowlegements:} I would like to thank J. Garriga and
 A. Kehagias for useful remarks.
We acknowledge partial support by
the European Commission RTN programme under 
contract HPNR-CT-2000-00131 and
by MCYT FPA 2001-3598 and CIRIT GC 2001SGR-00065.

%\begin{references}

%\end{references}


\begin{thebibliography}{4}

\bibitem{wmap}
C.L. Bennett {\it et al}., ``First Year Wilkinson Microwave Anisotropy
Probe (WMAP) Observations: Preliminary Maps and Basic Results,'' 
[astro-ph/0302207].


\bibitem{perl}
R.R. Caldwell, R. Dave and P.J. Steinhardt, Phys. Rev. Lett. {\bf
80}, 1582 (1998); 
%[astro-ph/9708069]; 
N. Bahcall, J.P. Ostriker, S.
Perlmutter and P.J. Steinhardt, Science {\bf 284}, 1481 (1999).
%[astro-ph/9906463].

\bibitem{exponencial}
Q. Shafi and C. Wetterich, Phys. Lett. {\bf B129}, 387 (1983); 
F. Lucchin and S. Matarrese, Phys. Rev. {\bf D32}, 1316 (1985); 
J.D. Barrow, A.B.
Burd and D. Lancaster, Class. Quant. Grav. {\bf 3}, 551 (1986);
J.J. Halliwell, Phys. Lett. B185 (1987) 341;
 A.~B.~Burd and J.~D.~Barrow,
%``Inflationary models with exponential potentials'',
Nucl.\ Phys.\ B {\bf 308}, 929 (1988).


%J.~D.~Barrow,
%``Slow Roll Inflation In Scalar - Tensor Theories,''
%Phys.\ Rev.\ D {\bf 51}, 2729 (1995).
%%CITATION = PHRVA,D51,2729;%%

\bibitem{kehagias}
A.~Kehagias and G.~Kofinas,
``Cosmology with exponential potentials,''
arXiv:gr-qc/0402059.
%%CITATION = GR-QC 0402059;%%

\bibitem{chimento}
L.~P.~Chimento,
 %``General Solution To Two-Scalar Field Cosmologies With Exponential
%Potentials,''
Class.\ Quant.\ Grav.\  {\bf 15}, 965 (1998).
%%CITATION = CQGRD,15,965;%%

\bibitem{townsend} P. Townsend, ``Cosmic acceleration and M-Theory", 
to appear in proceed. of ICMP2003, Lisbon, 
hep-th/0308149. 



\bibitem{chim2}
L.~P.~Chimento, A.~E.~Cossarini and N.~A.~Zuccala,
%``Isotropic and anisotropic N-dimensional cosmologies with exponential
%potentials,''
Class.\ Quant.\ Grav.\  {\bf 15}, 57 (1998).
%%CITATION = CQGRD,15,57;%%


\bibitem{town} P. Townsend and M. Wohlfarth, Phys.
Rev. Lett. {\bf 91}, 061302 (2003) [hep-th/0303097].

\bibitem{empa}
 R. Emparan and
J. Garriga, JHEP {\bf 0305}, 028 (2003) [hep-th/0304124].

\bibitem{recent}
L.~Cornalba and M.~S.~Costa,
Phys.\ Rev.\ D {\bf 66}, 066001 (2002)
[hep-th/0203031];
C-M. Chen, D.V. Gal'tsov and M. Gutperle, Phys. Rev. {\bf D66},
024043 (2002) [hep-th/0204071]; N. Ohta, Phys. Lett. {\bf B558},
213 (2003) [hep-th/0301095]; 
N. Ohta, Phys. Rev. Lett. {\bf 91}, 061303 (2003) [hep-th/0303238]; 
S. Roy, Phys. Lett. {\bf B567}, 322 (2003) [hep-th/0304084]; 
M.N.R. Wohlfarth, Phys. Lett. {\bf B563}, 1 (2003) [hep-th/0304089]; 
M. Gutperle, R. Kallosh and A. Linde,
JCAP {\bf 0307}, 001 (2003) [hep-th/0304225]; C-M. Chen, P-M. Ho,
I.P. Neupane, N. Ohta and J.E. Wang, JHEP {\bf 0310}, 058 (2003)
[hep-th/0306291];
I.~P.~Neupane, [hep-th/0311071].




\bibitem{russo}
J.~G.~Russo,
``Cosmological string models from Milne spaces and SL(2,Z) orbifold,''
Mod. Phys. Lett. A19, (2004)  421 [hep-th/0305032].
%%CITATION = HEP-TH 0305032;%%

 \end{thebibliography}
\end{document}